\documentclass[aps, prl, twocolumn, superscriptaddress]{revtex4-1}

\usepackage{bm}
\usepackage{graphicx}
\usepackage{epstopdf}
\usepackage{color}
\usepackage{soul}
\usepackage{cleveref}
\usepackage{amssymb,amsfonts,amsmath} 
\def\fig{Fig.}
\def\PENN{\small Department of Mechanical Engineering \& Applied Mechanics, University of Pennsylvania, Philadelphia, 19104, USA}
\def\NIST{\small Polymers \& Complex Fluids Group, National Institute of Standards and Technology, Gaithersburg, 20899, USA }

\begin{document} 

\title{Flow Resistance and Structures in Viscoelastic Channel Flows at Low Re}
\author{Boyang Qin}  
\affiliation{\PENN}  
\author{Paul F. Salipante}
\affiliation{\NIST}  
\author{Steven D. Hudson}
\affiliation{\NIST}  
\author{Paulo E. Arratia} 
\affiliation{\PENN}  

\begin{abstract}
The flow of viscoelastic fluids in channels and pipes remain poorly understood, particularly at low Reynolds numbers. Here, we investigate the flow of polymeric solutions in straight channels using pressure measurements and particle tracking. The law of flow resistance is established by measuring the flow friction factor $f_{\eta}$ versus flow rate. Two regimes are found: a transitional regime marked by rapid increase in drag, and a turbulent-like regime characterized by a sudden decrease in drag and a weak dependence on flow rate. Lagrangian trajectories show finite transverse modulations not seen in Newtonian fluids. These curvature perturbations far downstream can generate sufficient hoop stresses to sustain the flow instabilities in the parallel shear flow. 
\end{abstract}

\maketitle \thispagestyle{plain}

Fluids containing polymers are found in everyday life (e.g. foods and cosmetics) and in technology spanning the oil, pharmaceutical, and chemical industries. A marked characteristic of polymeric fluids is that they often exhibit non-Newtonian flow behavior such as viscoelasticity~\cite{1999LarsonBook,bird1987}. Mechanical (elastic) stresses in such fluids are history-dependent and develop with time scale $\lambda$, which is proportional to the time needed for a single polymer molecule to relax to its equilibrium state in dilute solutions. These stresses grow nonlinearly with shear rate and can dramatically change the flow behavior~\cite{1999LarsonBook,bird1987}. For example, the presence of of polymer in turbulent pipe flows can suppress eddies and lead to large reduction in flow friction~\cite{1975Virk,2008White}. At low Reynolds numbers (Re), where inertia is negligible, elastic stresses can lead to flow instabilities not found in ordinary fluids like water~\cite{1989Muller,1990Larson,1993McKinley,1998Groisman,2002Arora,2006Arratia,2007Poole,2013Pan}. They can also exhibit a new type of disordered flow -- elastic turbulence --  a turbulent-like regime existing far below the dissipation scale~\cite{2004Groisman, 2000Groisman,2001Groisman, 2010Fardin}. 

Recently, there has been mounting evidence that the flow of viscoelastic fluids in pipe and channel flows are nonlinearly unstable and undergo a subcritical instability at sufficiently high flow rates even at low Re~\cite{2004Meulenbroek, 2003Bertola,2005Morozov,2007Morozov,2011Bonn,2013Pan,2017Qin}. Theoretical investigations have used nonlinear perturbation analysis to show that a subcritical bifurcation can arise from linearly stable base states~\cite{2004Meulenbroek, 2005Morozov, 2007Morozov}, while non-modal stability analysis predicts transient growth of perturbation~\cite{2008Hoda,2010Jovanovic,2011Jovanovic}. Subsequent experiments in small pipes found unusually large velocity fluctuations that are activated at many time scales~\cite{2011Bonn} as well as hysteretic behavior ~\cite{2003Bertola}. More recently, experiments in a long microchannel using a linear array of cylinders as a way to perturb the (viscoelastic) flow showed an abrupt transition to irregular flow and that the velocity fluctuations are long-lived~\cite{2013Pan,2017Qin}. The unstable flow exhibits features of Newtonian turbulence such as power-law behavior in velocity spectra, intermittency flow statistics, and irregular structures in the streamwise velocity fluctuation \cite{2017Qin}. Taken together, these results show that polymeric solutions flowing in straight channels can undergo a subcritical transition -- a sudden onset of sustained velocity fluctuations above a perturbation threshold and a critical flow rate. This scenario is akin to the transition from laminar to turbulent flow of Newtonian fluids in pipe flows~\cite{1883Reynolds,2011Avila}. The main distinction is that the instability is caused by the nonlinear elastic stresses and not inertia. Unlike the Newtonian pipe turbulence, however, little is known about the basic structures organizing the instability and the law of resistance (i.e. pressure loss due to friction) as the flow transitions from a stable to an unstable state. 

\begin{figure}[h!]
\includegraphics[width=0.38\textwidth]{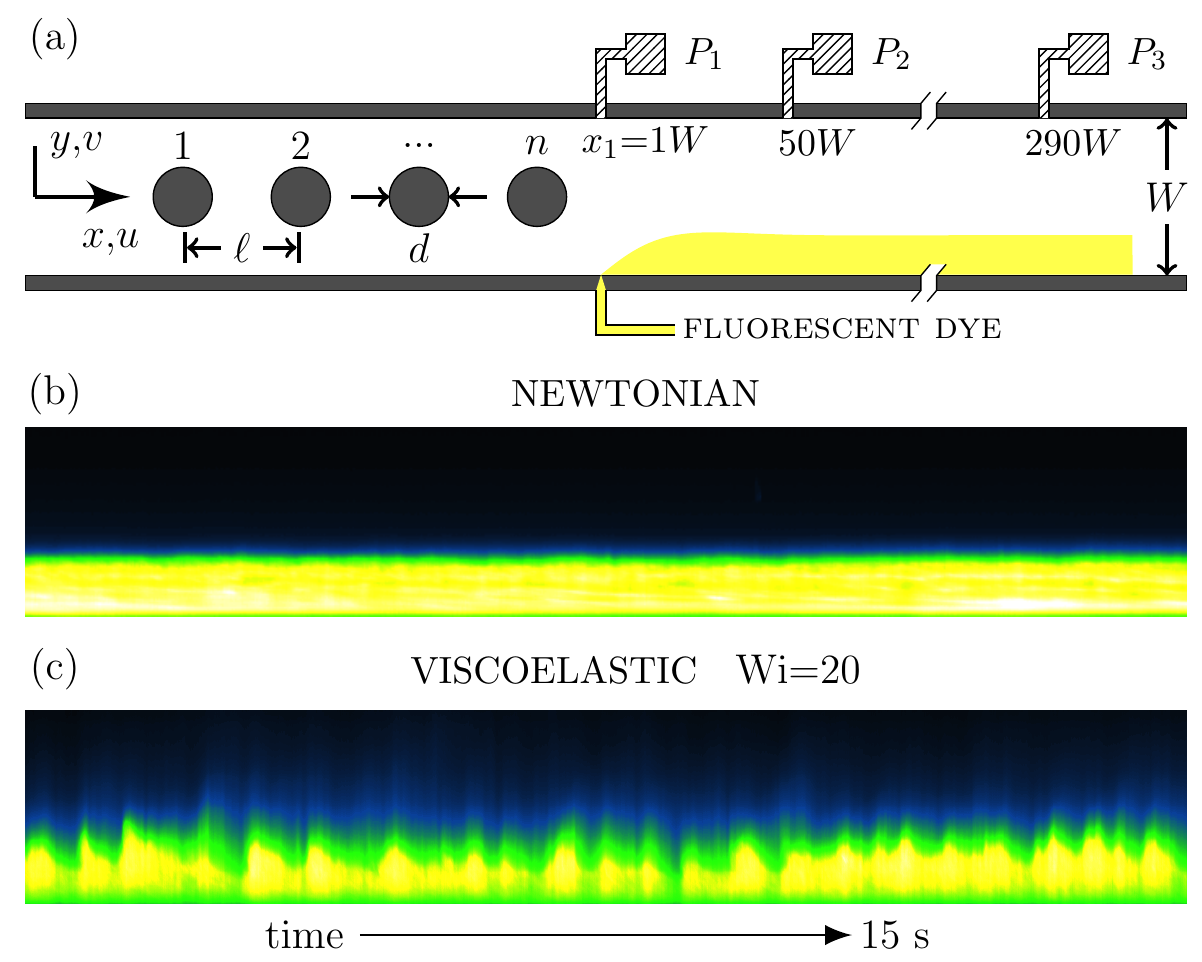}\vspace{-3mm}
\caption{\label{Fig1}(color online). (a) Schematic of the microchannel, showing location of pressure sensors and the dye injection scheme. (b,c) Space time dye patterns for $n$ = 15 and $x$ = 200$W$ in the parallel shear region, (c) viscoelastic fluid at Wi=20 and (b) Newtonian case at identical flow rate.
}
\end{figure}

In this manuscript, we investigate the flow of polymeric solutions in a straight micro-channel at low Re using pressure measurements and particle tracking methods. Pressure measurements show that the flow resistance increases relative to the stable viscoelastic base flow, following the transition from laminar to a ``turbulent-like" state, cf. \fig~1(c). This behavior is analogous to Newtonian turbulence where the friction factor increases as the flow transitions from laminar to turbulent except that here the governing parameter is the Weissenberg number ($Wi$), defined as the product of the fluid relaxation time $\lambda$ and the flow shear-rate $\dot{\gamma}$. The rise in flow resistance is related to enhanced elastic stresses and suggests flow patterns not seen in the (viscoelastic) laminar regime. We find that, far downstream from the initial perturbation, tracer particles follows wavy trajectories with spanwise modulation not found in the stable unperturbed flow (cf. \fig~5).  We believe that the increase in flow resistance is connected to the appearances of these wavy particle motions.  A new law of resistance for viscoelastic channel flows is proposed to capture this increase in drag. %

Experiments are conducted using a straight microchannel with equal width and depth ($W = D =$ 100 $\mu$m), fabricated using standard soft-lithography methods. The device schematic is shown in \fig~1(a). The channel length is much larger than its width $L/W$ = 330 and is divided into two regions. The first region consists of a linear array of fifteen cylinders ($n=15$) that extends for 30$W$. The diameter of the cylinders is $d=0.5W$ and their center to center separation is $\ell= 2W$; the last cylinder in the array is at $x = 0$. An unperturbed control case with no cylinders ($n=0$) is used as the linearly stable viscoelastic case. The second region follows the array of cylinders and consists of a long parallel shear flow 300$W$ in length. In order to measure pressure signals, sensors are placed at three locations in the parallel shear region, $x_1=1W, x_2=50W, x_3=290W$. 
Pressure signals are recorded for over 2 hours with a 5 ms resolution. The pressure drop per length between sensor 1 and 2 is $p_1(t)= (P_1-P_2)/(x_2-x_1)$ and similarly $p_2(t) = (P_2-P_3)/(x_3-x_2) $ for the segment between 2 and 3 (see \fig~1a).

The polymeric solution is prepared by adding 300 ppm of polyacrylamide (PAA) (18$\times$ $10^6$ MW) to a viscous Newtonian solvent (90\% by mass glycerol aqueous solution); the PAA polymer overlap concentration $c^*$ is 350 ppm \cite{2003Meadows} and $c/c^*$ = 0.86. This polymeric solution has a nearly constant viscosity of around $\eta$ = 300 mPa$\cdot$s (see SM \cite{Qin_SM} for fluid rheology). The Newtonian solvent, 90\% by mass glycerol in water, has constant viscosity of 220 mPa$\cdot$s and is also used for comparison. Throughout our experiment, the Reynolds number is kept below 0.01, where $Re = \rho U H/\eta$, $U$ is the mean centerline velocity, $H$ is the channel half-width, and $\rho$ is the fluid density. We characterize the strength of the elastic stresses compared to viscous stresses by the Weissenberg number \cite{1993Magda,1993McKinley}, defined here as $Wi(\dot \gamma ) = N_1(\dot\gamma )/2\dot\gamma \eta(\dot\gamma)$, where $\dot\gamma = U/H$ is the shear rate and $N_1$ is the first normal stress difference (see Supplemental Material \cite{Qin_SM} for details). 

We begin by investigating the flow patterns formed when a stream of fluorescent dye is injected one channel width ($1W$) after the last post. The dye spreading and patterns are then visualized far downstream in the parallel shear region, 200 channel widths downstream from the last post ($x = 200W$). Figure 1 show the spatio-temporal profile of the dye intensity along the device's cross section ($y$) for a channel containing 15 posts ($n=15$) for Newtonian (\fig~1b) and viscoelastic  (\fig~1c) fluids. For the Newtonian case, the profile shows typical laminar dye layer with minimal dye penetration into the undyed stream, except for diffusion. (Similar behavior is observed with viscoelastic fluids for the $n=0$ case.) A different dye pattern is observed when the Newtonian fluid is replaced by the polymeric solution under the same conditions. The viscoelastic case, shown in \fig~1(c) at $Wi\approx20$, shows irregular flow patterns with spikes of dye penetration into the undyed fluid stream. The observed fluctuations in time suggest the presence of velocity modulations in space. Indeed, we will show later that particle trajectories do exhibit wavy and coherent motions in the parallel shear region.


\begin{figure}[h!]
\includegraphics[width=0.3\textwidth]{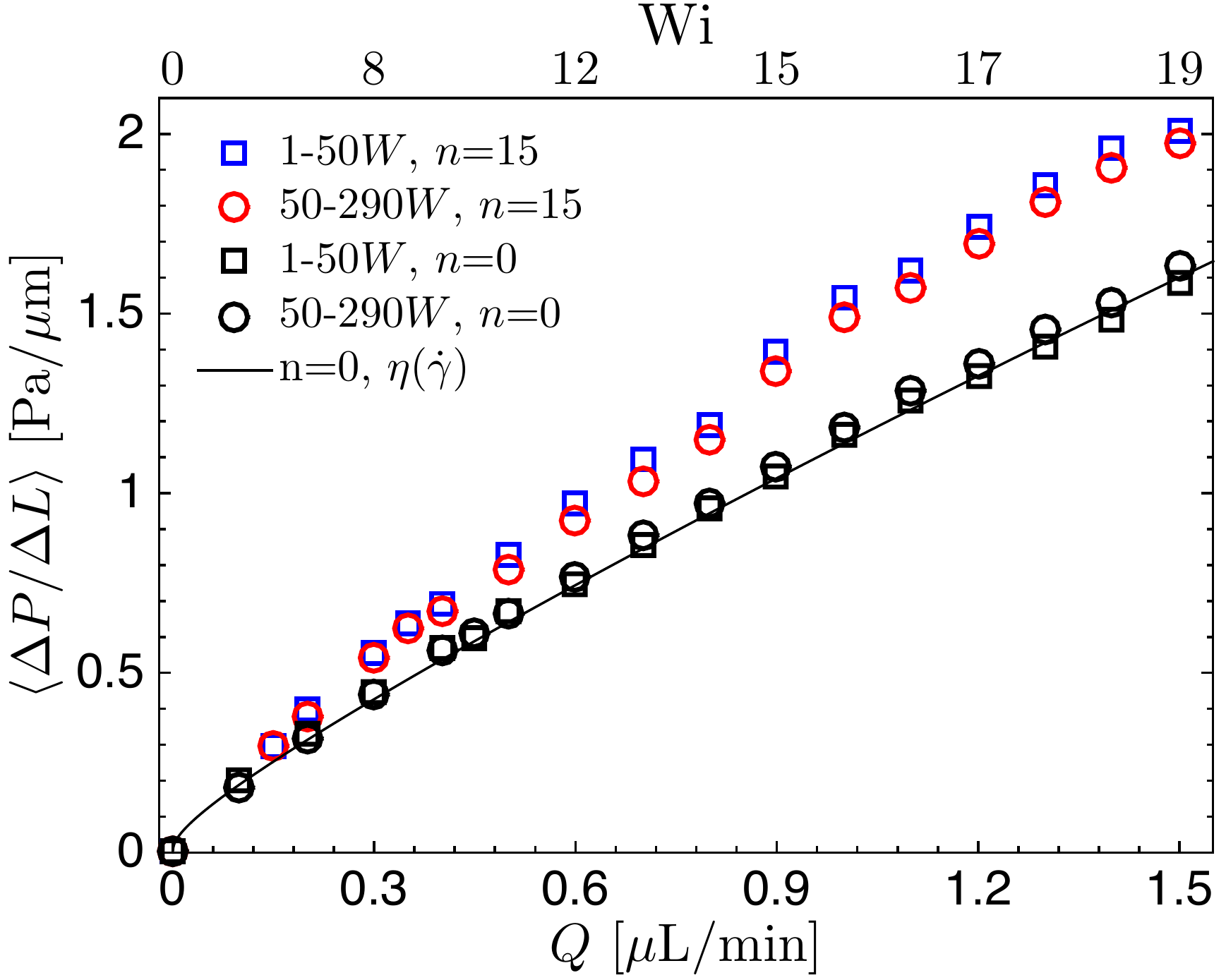} \vspace{-3mm}
\caption{\label{Fig2}(color online). Pressure drop per unit length as a function of flow rate $Q$ and $Wi$ for $n$=15 and $n$=0 cases. Solid line represents estimation using wall shear rates and viscosity from rheology measurements. Error bars are less than marker size and not shown here.
}
\end{figure}
As mentioned before, little is known about elastic turbulence in channel flows. Importantly, there is no known law of resistance for such flows. Here, we propose a new law of friction for polymeric solutions in channel and pipe flows. Pressure drop is measured along the parallel shear region using sensors that are placed at three locations, $x_1=1W, x_2=50W, x_3=290W$. 
The pressure drop per length signals $p_1, p_2$ are recorded for approximately $10^4$ seconds (with 5 ms resolution). Figure 2 shows the mean pressure drop values for viscoelastic fluids for $n=0$ and $n=15$ cases as a function of flow rate $Q$ and $Wi$. We note that the statistical mean of the reported signals measure the aggregate flow resistance encountered to sustain a constant mass flow rate. As expected, the pressure drop or flow resistance increases with flow rate and $Wi$. The pressure drop for the $n=0$ case slightly deviates from the Newtonian case (i.e. $\triangle P \sim Q$) due to mild shear-thinning in fluid viscosity. These effects can be accounted for by estimating the pressure drop using wall shear rate and corresponding viscosity $\eta(\dot\gamma)$ measured using a cone-and-plate rheometer, as shown by the solid line in \fig~2. No significant difference is found between $p_1$ and $p_2$ for $n=0$ case as expected, since entrance effects are minimized by using a tapered inlet that generates minor disturbance relative to that of the cylinder array. For $n=15$, we find a clear increase in pressure drop relative to the $n=0$ case; the two pressure segments $p_1$ and $p_2$ show little to no difference. This increase in flow resistance cannot be explained by solely shear-thinning effects and is related to the development of additional elastic stresses in the flow as the $Wi$ is increased. It also indicates that more energy is necessary to keep the same flow rate compared to a stable viscoelastic channel flow. 

\begin{figure}[h!]
\includegraphics[width=0.4\textwidth]{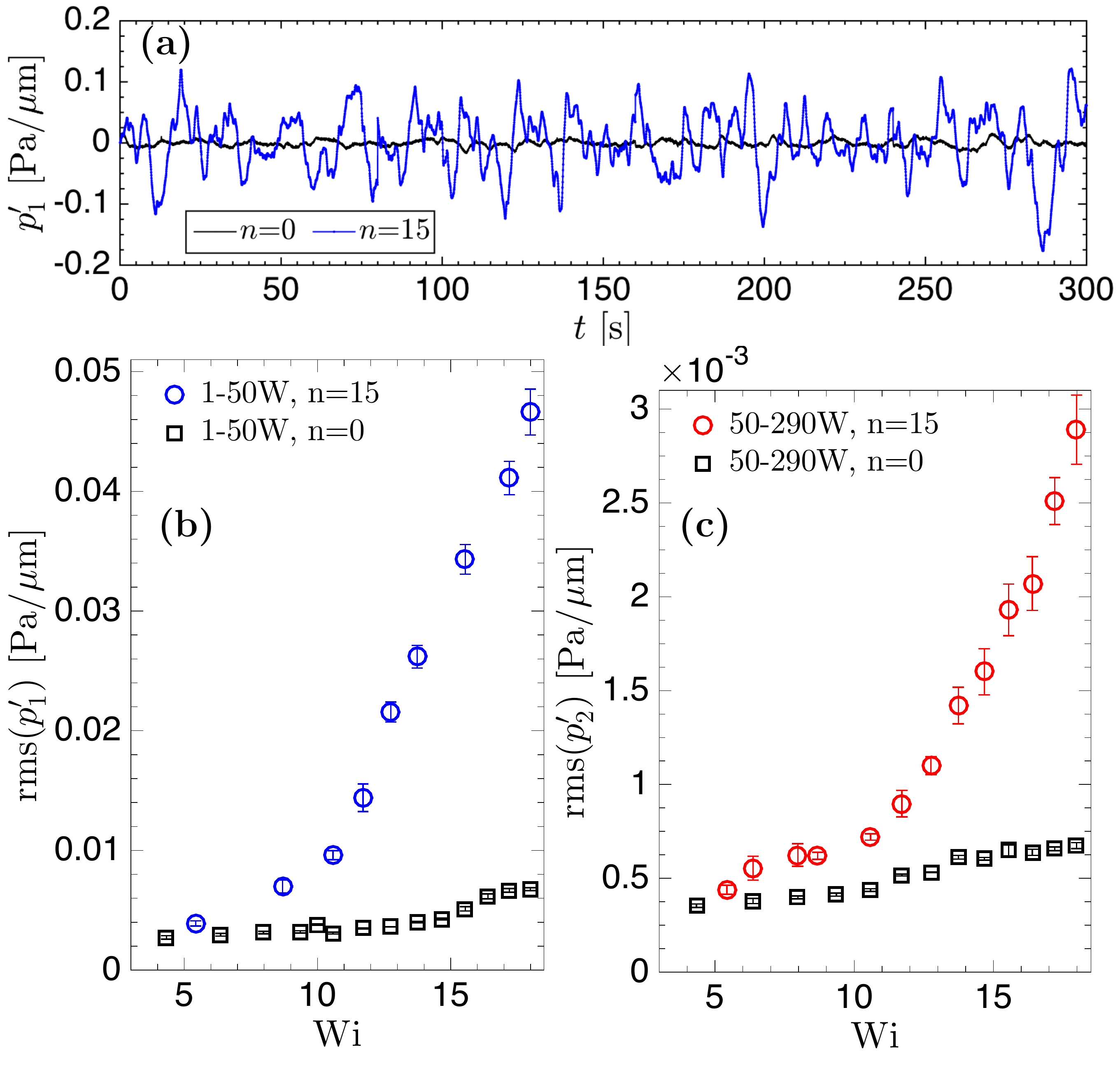}\vspace{-2.5mm}
\caption{\label{Fig3}(color online). (a) Pressure gradient fluctuations for $p_1'(t)$ between $x=1W$ and $50W$ for $n$=15 case, compared with the unperturbed $n$=0 case, $Wi=18$. (b,c) Root-mean-square (rms) of the pressure gradient fluctuations as a function of Wi for $n=0$ versus $n=15$, (b) $p_1'$ and (c)  $p_2'$.
}
\end{figure}

The increase in flow resistance is closely associated with the onset of pressure fluctuations (\fig~3). Figure 3(a) shows sample time records of pressure fluctuations $p_1'(t)$ for viscoelastic fluids at $Wi=18$ in devices with $n=0$ (black line) and $n=15$ (blue line). We observe a clear increase in the pressure fluctuations far downstream the cylinders once they are introduced in the flow. Figures 3(a) and 3(b) show root-mean-square (rms) values of the pressure fluctuations of the $p_1'$ and $p_2'$ segments, respectively, as a function of $Wi$ for the $n=15$ and $n=0$ cases. For the $n=0$ case, pressure fluctuations remain relatively small and steady, nearly independent of $Wi$; the small increase in pressure fluctuation at the higher values of $Wi$ may be due to entrance effects. We find that for both segments, $p_1'$ and $p_2'$, the rms values show significant departure from the stable $n=0$ case and a marked increased with increasing $Wi$. The values of the rms($p_1'$) and rms($p_2'$) start to depart from the $n=0$ trend at $Wi\approx5$ and grows weakly until $Wi\approx 9$. This is followed by a much steeper growth for $Wi \gtrsim 9$. This trend in pressure fluctuation measurements agrees relatively well with measurements of velocity fluctuations, for $n=15$ case, which established that the linear instability associated with the flow around the upstream cylinders occurs at $Wi\approx4$ and the onset of subcritical instability occurs at $Wi \approx 9$~\cite{2013Pan,2017Qin}.

Since pressure data is now available, one can investigate the law of (flow) resistance for viscoelastic channel flows as a function of $Wi$. This is analogous to measuring the Darcy friction factor for Newtonian pipe flows as a function of Re~\cite{1944Moody}. Traditionally, the friction factor $f$ is defined as $(\Delta P/\Delta L)/(\rho U^2/2W)$, where $W$ is the channel width and $U$ is the fluid mean velocity. As long as variations are small (e.g. smooth pipes), the friction factor $f$ is solely a function of $Re$ such that $f=f(Re)$. In what follows, we proposed an analogous law of resistance for viscoelastic channel flows as a function of $Wi$. As noted earlier, the values of $Re$ in our experiments are quite small ($Re \lesssim 10^{-3}$). Since fluid inertia is negligible, we propose to scale the pressure drop by the fluid shear stresses across the channel and define a viscous friction factor $f_\eta$ as  $[(\Delta P/\Delta L)/(c\eta_w \dot{\gamma}_w/W)]$, where $\dot{\gamma}_w$ is the wall shear-rate, $\eta_w$ is the correspondng viscosity and $c$ is a geometry factor ($c=4.06$ for square duct). 

\begin{figure}[h!]
\includegraphics[width=0.30\textwidth]{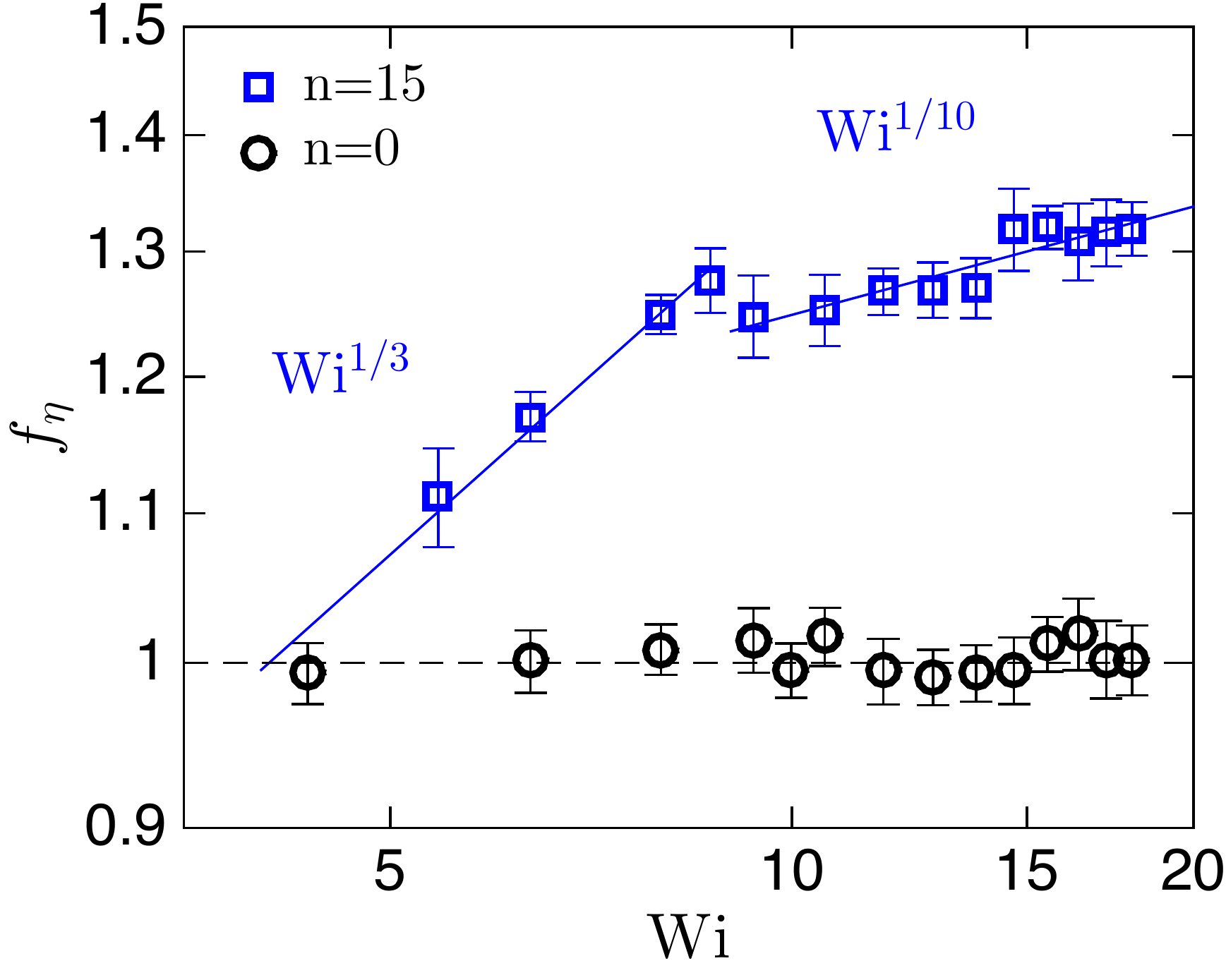}\vspace{-3mm}
\caption{\label{Fig4}(color online). Viscous wall friction factor $f_{\eta}$ (definition see text) as a function of Wi for $n=0$ and $n=15$.
}
\end{figure}

Figure 4 shows the values of $f_\eta$ as a function of $Wi$ for polymeric solutions in channels with $n=0$ and $n=15$. For $n=0$, we find that the $f_{\eta}$ is independent of $Wi$ indicating that flow resistance is purely governed for by viscous effects, which are well accounted for by the normalization. For $n=15$, on the other hand, we observe an increase in flow resistance $f_{\eta}$ as $Wi$ is increased and find that $f_{\eta} \sim Wi^{1/3}$ up to $Wi\approx9$. Surprisingly, we find a second regime for $Wi \gtrsim 9$ in which a sudden decrease in $f_{\eta}$ is observed followed by a weak dependence on $Wi$. This relative decrease in drag or friction seems to suggest the emergence of a new flow state which has yet to be explored in detail. The data shown in \fig~4 also suggests that the initial $f_{\eta} \sim Wi^{1/3}$ regime is associate with a transitional flow that is then followed by a turbulent-like state. Similar to Newtonian pipe flows, there is an initial increase in drag followed by a sudden decrease once the flow becomes turbulent. 

Next, we investigate the structure of the viscoelastic flow for n=15 and $Wi$=18; this is the regime in which we expect flow instabilities but quantifying the presence of flow structures has been difficult due to the weak spanwise velocity component relative to the mean shear \cite{2017Qin}. To interrogate the flow with enough spatial and temporal resolution, we use a novel three-dimensional holographic particle tracking method (hPTV)~\cite{2010Cheong,2017Salipante}. The flow is seeded with tracers (1~$\mu$m diam) which are imaged under microscope with high speed camera (5000 fps). Using a coherent light source, particle positions are reconstructed from the light scattering field projected onto the imaging plane (details see SM \cite{Qin_SM}). The uncertainty in particle centroid is approximately 30~nm for the in-plane ($x,y$) components and the seeding density is dilute (10$^{-5}$ volume fraction). The measurement window is located at $x$ = 200$W$ in the parallel shear region and extends for $2.5W$ streamwise and $0.9W$ spanwise.

\begin{figure}[h!]
\includegraphics[width=0.499\textwidth]{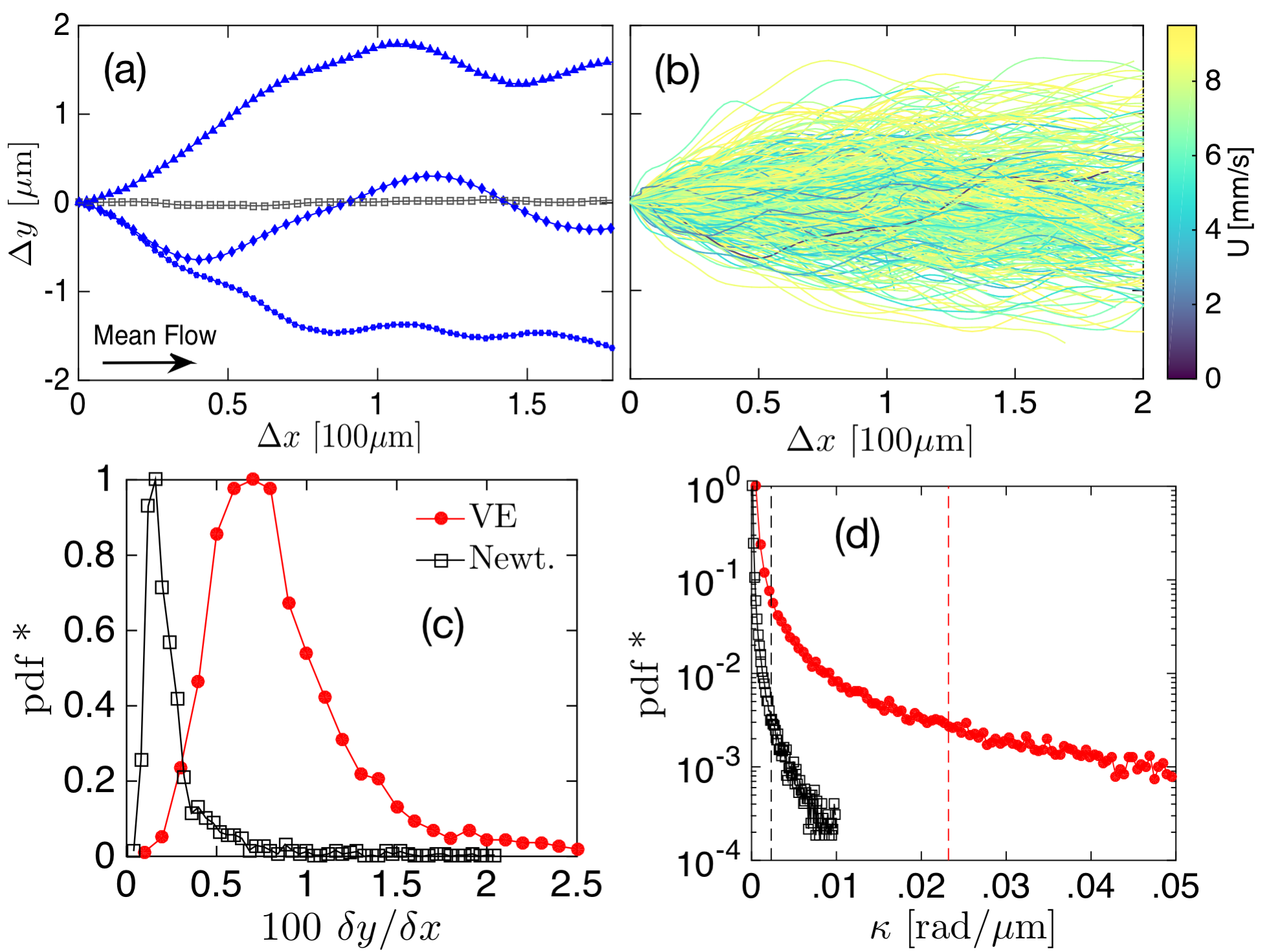}\vspace{-3mm}
\caption{\label{Fig5}(color online). (a) Particle trajectories in the streamwise ($x$) and spanwise ($y$) direction; blue lines represent the $n$=15 viscoelastic case at $Wi$=18 and the gray line is Newtonian at identical conditions. (b) Collection of trajectories colored by speed. Distributions of (c) cumulative transverse to streamwise displacements and (d) trajectory curvatures, where dash line represents population mean.}
\end{figure}

Figure 5(a) shows sample particle trajectories for the Newtonian (grey symbols) and viscoelastic (blue symbols) fluids for the $n=15$ case at $Wi=18$. While the particle trajectory in the Newtonian case is fairly rectilinear (following the mean flow direction) with little lateral movement, particle trajectories in the viscoelastic fluid case display a relatively pronounced waviness and lateral movement. This is not isolated to a few particles and \fig~5(b) shows the full extent of the spanwise spread of the Lagrangian trajectories for 2000 trajectories sampled uniformly in the channel. Such wavy structures underlie the irregular dye transport patterns seen in \fig~1(c). We quantify these deviations from the base-flow by calculating the probability distribution function (pdf) of the ratio between transverse to streamwise particle cumulative displacements (\fig~5c) defined as $\delta y/\delta x = \sum |dy_i| / \sum |dx_i|$, where $dy_i$ and $dx_i$ are particle displacements between frames.
The Newtonian data (black) show minimal transverse component and set the measurement noise level. Results show that particles in the viscoelastic fluid exhibit small but finite values of transverse (spanwise) velocity and a broader distribution of individual particle end-to-end displacement. These results indicate the presence of subtle flow structures in viscoelastic fluids in parallel shear flows. While these deviations from the base-flow are small in absolute terms (2\% of the streamwise component), even small deviations in the velocity fields in viscoelastic fluids can represent significant increase in elastic stresses due to the nonlinear relationship between stress and velocity~\cite{2001Alves,2011Thomases}. 

Can these curved particle trajectories drive or maintain flow instabilities far downstream ($200W$)? Figure 5(d) shows the distribution of particle pathline curvatures at $200W$ for $Wi=18$, $n=15$. The trajectories have a mean curvature of $1/\mathcal{R} \approx .023$ $\mu$m$^{-1}$, which is an order of magnitude larger than the Newtonian counterpart. Using $N_1$ data (see SM \cite{Qin_SM}), we compute the Pakdel-McKinley condition defined as $[(\lambda U/\mathcal{R})Wi]^{1/2}$ \cite{1996Pakdel}. We find a value of approximately 7, which is sufficiently large to trigger flow instabilities. Similarly, we find that hoop stresses $N_{1}/\mathcal{R} = 8$~Pa/$\mu$m can be of the same order (or higher) than pressure drop $\Delta P/\Delta L = 2$~Pa/$\mu$m. These results strongly suggest that even small streamline curvature from velocity fluctuations can generate elastic stresses, which can sustain flow instabilities far downstream. 



In summary, we investigated the flow of viscoelastic fluids in a long, straight microchannel at low Re. This flow becomes unstable via a nonlinear subcritical instability at a critical $Wi$ for finite amplitude perturbations~\cite{2013Pan}. Pressure measurements are used to establish the \textit{law of resistance} for this flow (\fig~4). We find two regimes: (i) a transitional regime ($5 \lesssim Wi \lesssim 9$) in which the (viscous) friction factor $f_{\eta} \sim Wi^{1/3}$, and (ii) a turbulent-like regime ($Wi \lesssim 9$) in which a sudden reduction of $f_{\eta}$ is observed followed by a weaker dependence on flow rate that leads to $f_{\eta} \sim Wi^{0.1}$. This behavior is analogous to Newtonian pipe flows in which a sudden increase in drag is followed by a weaker dependence on Re. Dye and particle tracking data show the presence of weak flow structures far downstream in the parallel shear region ($200W$). In particular, we find small but finite particle lateral (spanwise) movement and transverse modulations relative to the Newtonian case (\fig~5). These particle trajectories have enough curvature and speed to generate hoop stresses that can sustain flow instabilities. Our results provide strong evidence for the ``instability upon an instability'' mechanism proposed for the finite amplitude transition of viscoelastic fluids in parallel flows~\cite{2005Morozov,2007Morozov} and provide new insights into the flow of polymeric solutions in channels and pipes. Even small perturbations in the velocity field can lead to significant changes in pressure drop and drag. 

We thank B. Thomases, A. Morozov, R. Poole, and M. Graham for fruitful discussions. P.E.A. acknowledges support from NSF CBET-1336171 and S.D.H. thanks NIST on a Chip funding.

\bibliography{referenceLR_v1}

\end{document}